\documentstyle[aps,multicol,epsf,epsfig]{revtex}
\newcommand{\be}{\begin{equation}}
\newcommand{\ee}{\end{equation}}
\newcommand{\bea}{\begin{eqnarray}}
\newcommand{\eea}{\end{eqnarray}}
\def\sig{{\boldmath$ \sigma$}}

\begin{document}

\title{Geometrically Reduced Number of  Protein Ground  State Candidates}
\author{ M.R. Ejtehadi$^1$\footnote{\it e-mail: reza@theory.ipm.ac.ir},
N. Hamedani$^{1,2}$ and V. Shahrezaei$^{1,2}$}
\address{$^1$ {\it Institute for studies in Theoretical Physics and Mathematics,
 Tehran  P.O. Box 19395-5531, Iran.}\\
$^2$ {\it Department of Physics, Sharif University of Technology,
 Tehran P.O. Box: 11365-9161, Iran.}\\} 

\maketitle
\begin{abstract}
Geometrical  properties of protein  ground states are  studied using an 
algebraic approach. It is shown that independent from inter-monomer
interactions, the collection of ground state candidates for any folded protein 
is unexpectedly small: For the case of a two-parameter Hydrophobic-Polar
lattice  model for $L$-mers, the number of these candidates  grows only 
as $L^2$. Moreover, the space of the interaction parameters of the model
breaks up into well-defined domains, each corresponding to one ground 
state candidate, which are separated by sharp boundaries. 
In addition, by exact enumeration, we show  there are some sequences which 
have  one absolute unique native state. 
These absolute ground states have perfect stability against change
of inter-monomer interaction potential.
\end{abstract}

\begin{multicols}{2}
%%%%%%%%%%%%%%%%%%%%%%%%%%%%%%%%%%%%%%%

It is well known that the biological functionality of  proteins  depends on the 
shape of their native states. This native structure is the unique 
 minimum free energy structure for the protein sequence
 \cite{Anfinsen}. Thus, the information about  effective inter-monomer 
interaction energy  and coding of the amino acids in the  sequence  is 
sufficient to determine the native structure.
There are many questions about the folding 
mechanism, stability, sensitivity to inter-monomer interactions and 
geometrical properties of these native states.
This has motivated extensive studies in the subject of the properties 
of native states in recent years. 

Basically, to determine the native states of a protein one needs to solve the 
problem with standard quantum mechanical calculations, however the complexity
of these macromolecules renders this impossible. A feasible 
approach to this problem is based on a  coarse-grained view to proteins. 
The most important point in this approach is  the choice of effective 
interactions between the monomers \cite{MJ}.
In this approach, all the necessary information about the protein's 
structure is encoded in
a binary contact matrix ($M$). The non-zero elements of this matrix 
correspond
to non-sequential neighbor monomers in the configuration \cite{Lifson}.
However, the shapes of the native states of sequences depend on 
the inter-monomer interactions \cite{Grosberg}, but there are some 
geometrical 
properties which distinguish the native state from other configurations 
\cite{Li,Vendr,self1,self2}.

The number of possible configurations for an $L$-mer 
is equal to the number of self avoiding walks
with $L-1$ steps. Since many of these walks give the same contact 
matrix, the number of possible contact matrices are much smaller 
\cite{Vendr}, although it is still very large, and grows exponentially with 
the length of protein.
There have been  some attempts  to reduce the number of possible protein 
configurations by considering  the compact structure space \cite{Dill},
or the minimum energy compact 
structure space \cite{Camacho}.
In the present work we look at the geometric constraints of native states 
in more detail. We show that there are some necessary geometrical 
properties for a state to be the ground state of a sequence.
This  leaves only a few candidates for the ground state of any sequence.
To find the ground state candidates for any sequence, one  need not
know anything about inter-monomer interactions, instead these candidates 
can be found by a simple comparison of the contact matrices.
One can then find the ground state from among these candidates by taking 
the inter-monomer interactions into account.
By limiting the number of ground state candidates, this method introduces 
a stability against the variation of interaction parameters and vastly 
reduces the computer time needed to find the native state for a 
particular sequence as one need not search through a huge number of 
configuration.

Without loss of generality, we use a hydrophobic-polar (HP) lattice model
\cite{Chan} in this paper. The argument can be  generalized to  any 
model with any number of monomer types with short range interactions.
The general form of interactions between $H$ and $P$ monomers in an HP model 
can be written as follows \cite{self1,Li1}:
\bea
\label{e1}
E_{HH} &=& -2 -\gamma -E_c, \nonumber\\
E_{HP} &=& -1 - E_c, \nonumber \\
E_{PP} &=& - E_c.
\eea
These potential energies are only between the non-sequential nearest
neighbours. Here $\gamma$ and $E_c$ are the mixing and compactness potentials 
respectively, two free parameters which are determined from experimental 
data. 
The compactness of the native states \cite{Dill} together with some 
physical arguments about inter-monomer interactions such as \cite{Li2}:
\bea
\label{e2}
E_{HH} < E_{HP} < E_{PP}, \nonumber \\
E_{HH} + E_{PP} < 2E_{HP},
\eea
restrict $\gamma$ and $E_c$ to positive values ($\gamma, E_c>0$), 
however, we need not consider such restrictions in our arguments.
At the first sight it might seem possible to arrive at any native state 
for a given sequence by changing $\gamma$ and $E_c$, but when we 
consider the geometrical properties of the ground state, we will find 
that these parameters are not powerful enough to select any 
configuration as the native state and native states are stable against 
the change of interaction parameters, in fact, universal solutions can 
be found for native states.

As explained in our previous works \cite{self1,self2}
if we consider $H=-1$ for hydrophobic monomers
 and $P=0$ for polar monomers, a given sequence can then  be represented 
by a binary vector (\sig). The energy of this sequence in a 
configuration characterized by a contact matrix $M$, can be written as:
\be
\label{e3}
E= -m -a\gamma- b E_c,
\ee
where $m$, $a$ and $b$ are three integers, related to
\sig and $M$ as follows:
\bea
\label{e4}
m&=& - {\mbox {\sig}}^{t} \cdot M \cdot{\bf 1}, \nonumber \\
a&=&  {1\over2} {\mbox{\sig}}^{t} \cdot M
 \cdot{\mbox{\sig}}, \nonumber \\
b&=& {1\over2} {\bf 1}^{t} \cdot M \cdot{\bf 1}.
\eea
Therfore, $m$ is equal to the number of all non-sequential neighbors 
of H monomers in the configuration,
$a$ is the number of  H-H contacts and $b$ is the number of all contacts.
It can be seen easily that the 
 following inequalities hold.
\be 
\label{e5}
m-b\le a\le {m\over 2} \le b.
\ee

Equation \ref{e3} suggests that the energy levels of a given sequence 
can be described by three integer numbers  $(m,a,b)$.
It is highly probable that these states are degenerate.
There are three kinds of  degeneracy:
(Type 1) $M=M'$ in which case two or more configurations
with different shapes  have the same contact matrix.
These configurations will remain  degenerate for any sequence,
and any choice of $\gamma$ and $E_c$. This type of degeneracy 
is more probable for configurations with low compactness.
note that the  configurations which are related to each 
other by spatial symmetries  i.e. rotation, reflection, etc., are the 
same and are not considered as separate. (Type 2) $(m,a,b)=(m',a',b')$ 
but $M \ne M'$; in this case one particular sequence has the same $m$,
$a$ and $b$ values in two or more configurations. This degeneracy persist 
for any value of $\gamma$ and $E_c$, but may dissappear for another 
sequence. Although this degeneracy depends on sequence coding,
but the  $b=b'$ condition is purely geometrical, and is a necessary 
condition for this degeneracy. (Type 3) $E=E'$, but $(m,a,b) \ne (m',a',b')$;
one sequence has the same energy in two different states $(m,a,b)$ and 
$(m',a',b')$, provided $\gamma$ and $E_c$ obey the following relation: 
\be
\label{e6}
(m-m') + (a-a')\gamma + (b-b') E_c = 0.
\ee
This degeneracy is related to both sequence coding \sig and inter-monomer 
interactions.

The first type of  these degeneracies is completely geometric. 
The second one depends on both geometry and sequence arrangement. These 
two types  don't depend on  the values of the interaction energies.
Thus, in the energy spectrum of any sequence there are some states 
which, independent of the potential,  are degenerate. If the ground state of 
one sequence is one of  these degenerate states, it means that this sequence
has not a unique native structure.
The third type  is not actually a degeneracy at all.
Equation \ref{e6} corresponds a 
line in  the parameter space of $E_c$ and $\gamma$. This line is a level 
crossing line. Degeneracy  occurs only on the line, and it needs highly 
accurate fine tuning. For the two sets of interaction energy parameters
on the two sides of this line, the  energy ordering 
of states is different. For any pair of  states there is such an ordering 
line. By drawing all ordering lines in the space of $E_c$ and $\gamma$, 
this space is divided into many ordering zones. We are only interested in 
the ground state, which means that many of these ordering lines are not 
relevant. Some of them only govern the ordering of excited states.
By removing the irrelevant lines, one gets a diagram which shows the ground 
state cells (Fig. 1). 
As mentioned before changing the inter-monomer interaction parameters
inside any of these cells does not change the ground state.
Mourik {\it et al.}  \cite{Mourik} introduced this picture
to  show  stability of native states against the interaction parameters 
\cite{note}. They only  looked at one of these cells in the neighbourhood of 
selected interaction values.
But by looking at the whole energy space, one can find all possible ground 
states and their corresponding cells.
Any such cell in the space of energy parameters associates with one 
ground  state candidate. The number of cells is  the number of 
candidates ($G_c($\sig$)$) for ground state. By drawing such diagram, 
one can easily find the ground state for any choice of $E_c$ and $\gamma$.
Fig. 1 shows this diagram for an  18-mer, which is the result of an 
exact enumeration of a two dimensional folding problem. 
The interesting point is that the number of ground state candidates is 
very small. In this example there are only five possible ground states.   
The cells marked with the numbers ``1" and ``2" correspond to type 1 
and 2 degenerate states respectively, therefore there is no unique 
native structure for these cells. 
The sequence in this example has two nondegenerate states.
These structures are shown in the figure. It is possible 
that all the ground  state candidates  of a given sequence  are degenerate. 
These sequences constitute universally bad sequences for any value of 
interaction parameters. It means that they do not have a native structure.

In Fig. 2 the histogram of $G_c($\sig$)$ for all $2^{18}$ sequences is shown.
The narrow line in this figure shows the result for all $2^{18}$ sequences,
and the thick line shows the remaining sequences after removing the 
bad ones. 
The interesting point in this diagram is the smallness of the mean value of
$G_c($\sig$)$, ($1.49$ for all sequences and $1.7$ for good sequences).
The maximum of $G_c($\sig$)$ for 18-mers is 6, and only 2 
sequences with length 18 have this maximum.
Comparison of $G_c($\sig$)$ for these sequences with the number of all
configurations ($\sim 10^7$), shows that the geometric constraints play
an important role in choosing a state as the ground state.
As this diagram shows, there are some sequences which regardless of the 
values for energy parameters, have only one unique ground state.
Fig. 3 shows one of these sequences and its unique native structure.
Indeed the native state of these sequences have perfect stability 
with respect to 
energy parameters. This enumeration shows that nearly  $17.8\%$ of the 
$2^{18}$ possible sequences have perfect stability  and absolute unique 
native states. Interestingly, our enumeration shows that these absolute 
native structures are between the most compact structures. 
Although the ratio of the perfectly stable proteins to all possible  
proteins decreases with increasing $L$, their actual number increases 
\cite{self3}. 
This suggests that for the proteins with typical length near natural 
proteins there are a small but non-zero fraction of perfectly stable 
sequences. The existence of these sequences may answer some  questions 
about protein folding. Their number is small compared with  the huge 
number of the possible amino-acids sequences, their native states are 
highly compact and are stable against the changes in the inter-monomer 
interactions.

The reason that there are few  ground state  candidates for
any sequence can be described by a geometrical argument.
Consider a three dimensional space with axes $X$, $Y$ and $Z$. Any state 
is represented by a point with coordinates $a$, $b$ and $m$ in this 
space (Fig. 4). All the states will be inside a pyramid according to 
equation \ref{e5}. For any value of $\gamma$ and $E_c$, let's consider
the following plane  perpendicular to the vector $(\gamma, E_c, 1)$:
\be
\label{e7}
z=-\gamma x -E_c y +(z_0 + \gamma x_0 +E_c y_0). 
\ee
If this plane contains the point $(x_0,y_0,z_0)=(a,b,m)$, 
the $z$ value on the $Z$ axis will be equal to $-E$. Thus, to find the 
ground state it is enough to move this plane from above until it touches a
state. This state is the ground state.
From this picture it is obvious that the possible ground states are in the
corners of the convex hull of the set of points (i.e. the polyhedral 
envelope of the states). If  $\gamma$
and $E_c$ can become  negative, all the corner points which 
can be seen  from the top view of this polyhedron, 
are  ground state candidates, but clearly that for  positive values of 
$\gamma$ and $E_c$ the number of possible ground states is even smaller. The 
cross section of the pyramid with a horizontal plane is a rectangle. For 
positive values of $E_c$ and $\gamma$  there is an  upper limit for 
possible ground sates.
It is equal to the number of possible states in the biggest horizontal 
rectangular cross section of the pyramid.
However the number of configurations grows as $z_{\rm eff}^L$,
where $z_{\rm eff}$ is effective coordination number  \cite{Fisher}, but 
the number of  ground state candidates grows very slower.
The maximum number  of contacts is of the  order of the  length of the
sequence, i.e.  $b_{\rm Max} \sim L$. Thus this upper limit grows as $L^2$.
This shows that the number of  ground state candidates grows much more slowly
than the number of configurations. For example, for 18-mers $b_{\rm 
Max}=10$,  the biggest 
cross section is a $6\times6$ rectangle. It thus gives $36$ as the maximum 
number of  ground state candidates. 
As can be seen in 
Fig. 2, the maximum number of  ground state candidates for 18-mers is 6 
according to exact enumeration, which is still much  smaller than the above 
estimate.

The dimension of state space is related to the  model and the number of 
energy parameters. For example,
If we look for the ground state in the space of compact configurations, 
$E_c$ is an irrelevant parameter. In this special case 
the space of energy parameters is  one dimensional (only $\gamma$), and the 
space of states is two dimensional ($a$ and $m$) \cite{self1,self2}.
This argument can be generalized to models with more than two kinds of 
monomers, and also to off-lattice models. For off-lattice models, it is 
necessary for the  energy function between monomers to be in the form of 
a step potential. For this form, a contact matrix can give the 
configuration energy by a relation similar to equation \ref{e3}.
 If the inter-monomer interaction has $t$
free parameters,  the energy levels can be described by  $t+1$ integer. 
The introduction of $n$-body interaction ($n>2$) only increases this 
difference the dimensionalities of state space and the space of energy 
parameters \cite{self4}.
Therfore, quite generally, the ground state candidates of any given 
sequence are between the corner states of  a hyper polyhedron in a hyper 
space which is very smaller than the number of all possible structures.

{\bf   Acknowledgements:} We would like to thank H. Seyed-Allaei, S. E. Faez, 
R. Gerami for useful discussions, R. Golestanian for 
helpful comments  on the style of presentation and N. Heydari for carefully 
reading the manuscript.
%%%%%%%%%%%%%%%%%%%%%%%%%%%%%%%%%%%%%%
\newcommand{\PNAS}[1]{ Pros.\ Natl.\ Acad.\ Sci.\ USA\ {\bf #1}}
\newcommand{\JCP}[1]{ J.\ Chem.\ Phys.\ {\bf #1}}
\newcommand{\PRL}[1]{ Phys.\ Rev.\ Lett.\ {\bf #1}}
\newcommand{\PRE}[1]{ Phys.\ Rev.\ E\ {\bf #1}}
\newcommand{\JPA}[1]{ J.\ Phys.\ A\ {\bf #1}}

\end{multicols}
%%%%%%%%%%%%%%%%%%%%%%%%%%%%%%%%%%%%%%%%%%%%%%%%%%%%%%%%%%%%%%%%%%%%%%%%%%%
\newpage
\begin{center}
\Large
Figure Captions \\[15mm]
\normalsize
\end{center}

{\bf Figure 1.}

\hspace{3cm}
\parbox{12cm}{
	The space of energy parameters for one particular sequence which is
	shown in the top of the picture is divided to five cells.
	The integer numbers $(m,a,b)$, inside any cell indicate the ground state 
	corresponding to the cells. Three of  these states are degenerate.
	The types of degeneracies for degenerate states and shape of structures
	for non-degenerates are indicated in the cells.}

\vspace{5mm}

{\bf Figure 2.}

\hspace{3cm}
\parbox{12cm}{ 
	The histogram of the number of ground state candidates for 18-mers
	with positive values of $\gamma$ and $E_c$ .
	The narrow and thick lines show the results for all sequences and  
	good sequences respectively. There are some ``good sequences" with only
	one ground state candidate.}

\vspace{5mm}

{\bf Figure 3.}

\hspace{3cm}
\parbox{12cm}{
	One perfectly stable sequence and its absolute native structure.
	For any positive value of $\gamma$ and $E_c$ this sequence is folded
	uniquely in the shown structure. }

\vspace{5mm}

{\bf Figure 4.}

\hspace{3cm}
\parbox{12cm}{
	State space of the particular sequence which is shown in figure 1.
	All states are inside a  diamond like polygon inside a pyramid.
	Top viewed corner points of this polygon are the ground state candidates.}

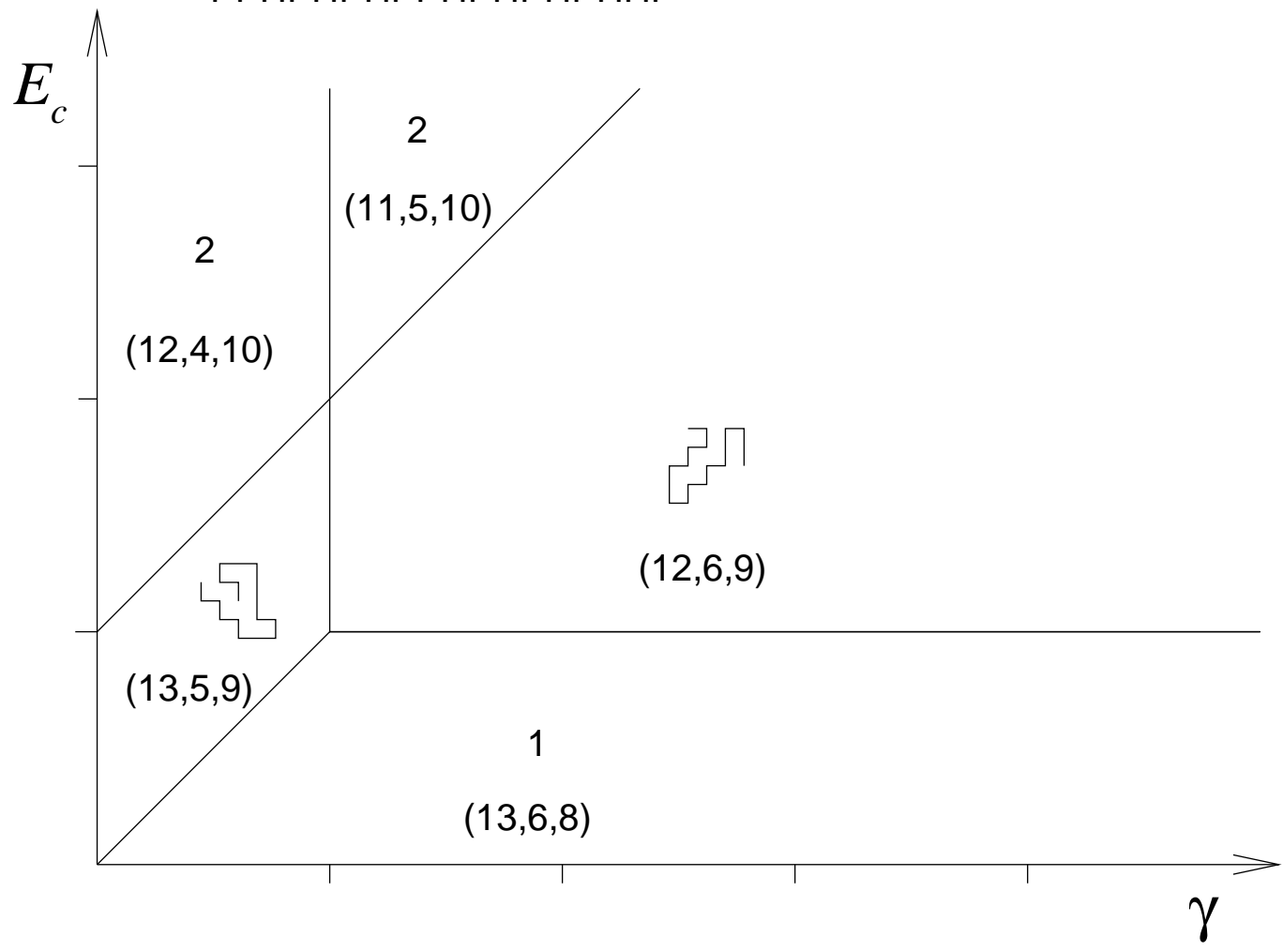
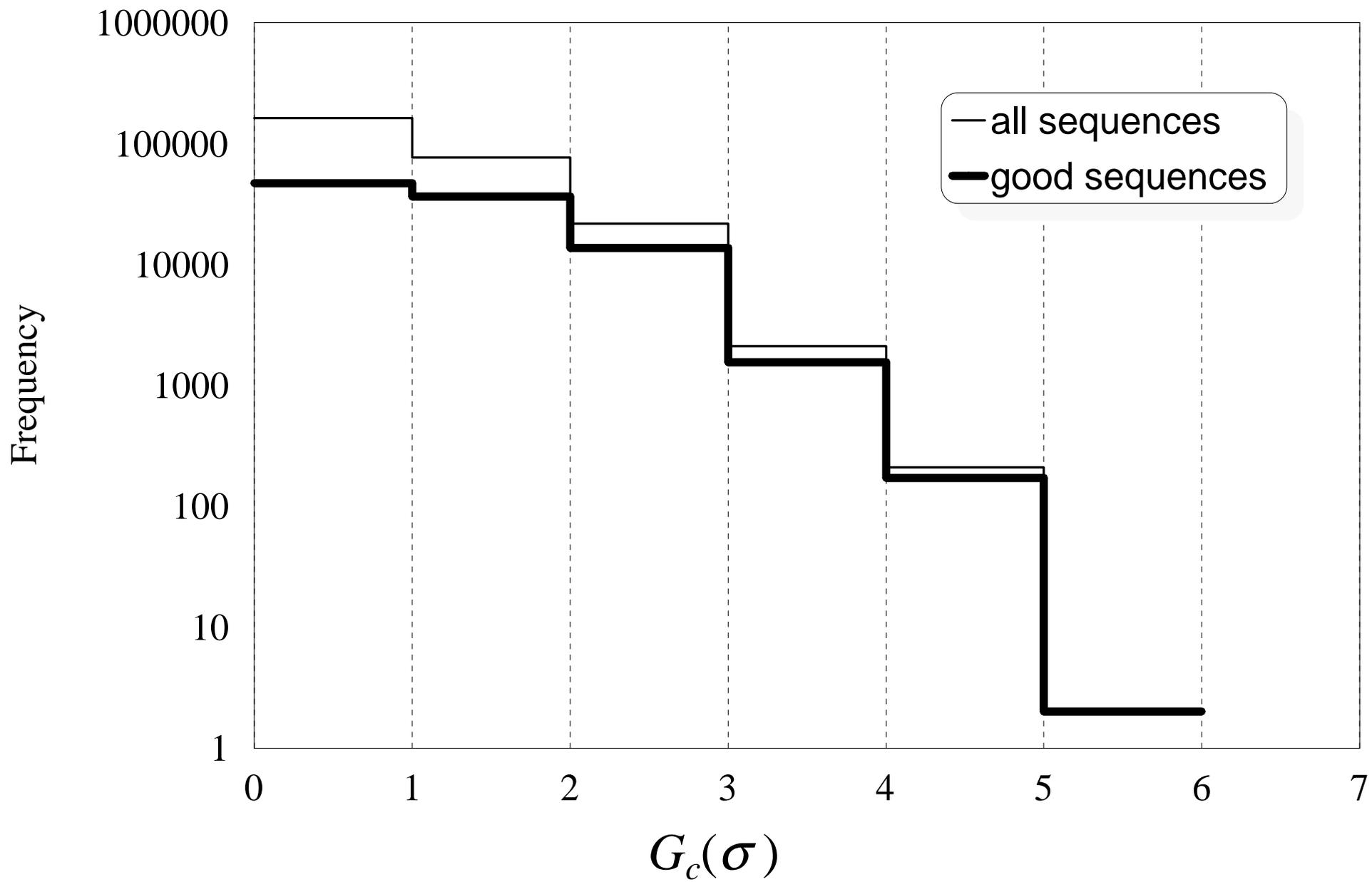
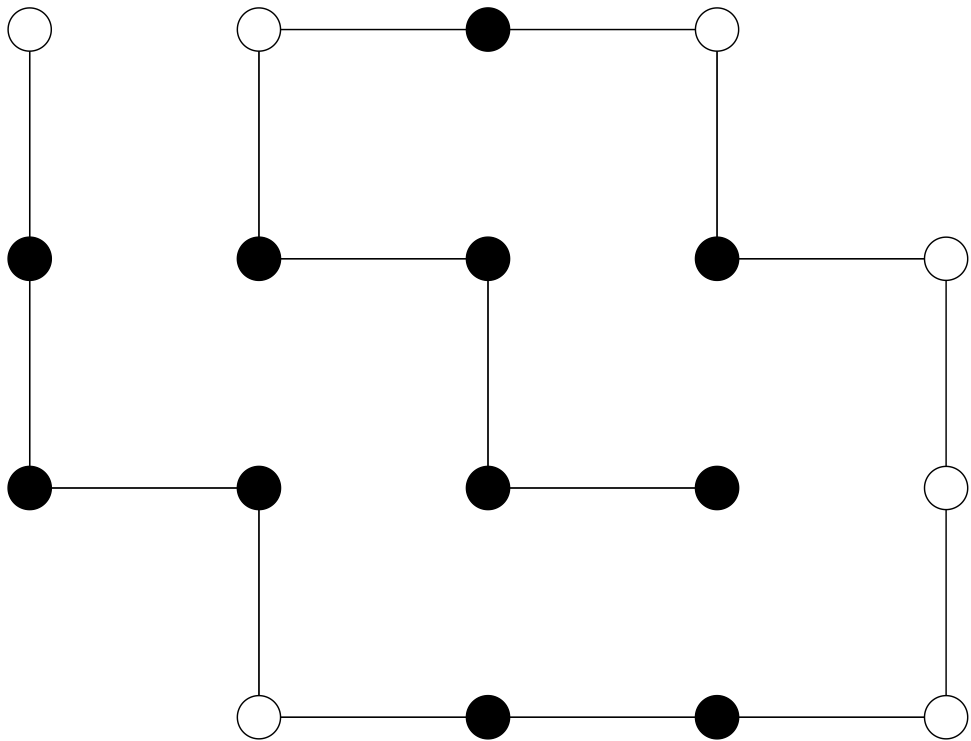
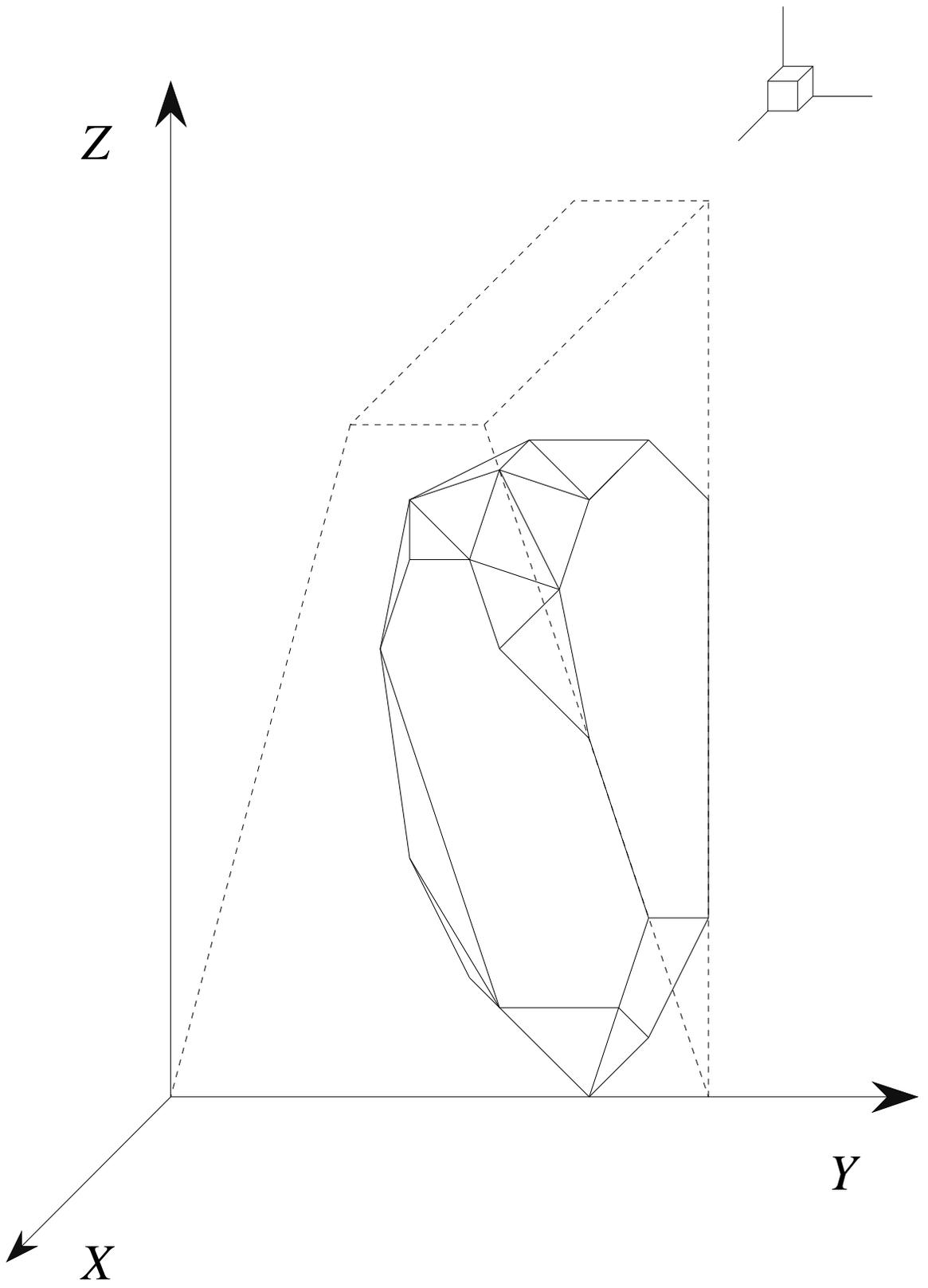
\end{document}